\documentclass[aps,prb,showpacs,byrevtex,showpacs,amsmath,amssymb,twocolumn,superscriptaddress]{revtex4}
\usepackage{epsfig}
\usepackage{color}

\newcommand{\be}{\begin{equation}}
\newcommand{\ee}{\end{equation}}
\newcommand{\bea}{\begin{eqnarray}}
\newcommand{\eea}{\end{eqnarray}}

\begin{document}
\bibliographystyle{prsty}
\title{Solving lattice density functionals  close to
  the Mott regime 
}

\author{  Zu-Jian Ying}
\affiliation{ Istituto dei Sistemi Complessi CNR and
Universit$\grave{a}$ di Roma ``La Sapienza", P. le A. Moro 2,
I-00185 Rome, Italy. }
\affiliation{ Beijing Computational Science Research Center, Beijing 100084, China. }
\author{Valentina Brosco}
\affiliation{ Istituto dei Sistemi Complessi CNR and
Universit$\grave{a}$ di Roma ``La Sapienza", P. le A. Moro 2,
I-00185 Rome, Italy. }
\author{Jos\'e Lorenzana}
\affiliation{ Istituto dei Sistemi Complessi CNR and
Universit$\grave{a}$ di Roma ``La Sapienza", P. le A. Moro 2,
I-00185 Rome, Italy. }
\thanks{jose.lorenzana@roma1.infn.it}
\begin{abstract}
We study a lattice version of the local density approximation (LDA) based on
Behte ansatz (BALDA). 
Contrary to what happens in Density Functional Theory (DFT) in the
continuum and despite its name, BALDA  displays some very non-local features
and it has a discontinuous functional derivative. The
same features  prevent the convergence of the self-consistent
Kohn-Sham cycle thus  hindering the study of BALDA solutions close to a Mott
phase or in the Coulomb blockade regime. 
Here we propose a numerical approach which,  differently from previous works, 
does not introduce {\sl ad hoc} parameters to smear out the singularity.
Our results are relevant for all lattice models where BALDA is
applied ranging from Kondo  systems to harmonically trapped Hubbard fermions.
As an example we apply the method to the study of a
one-dimensional lattice model with Hubbard interaction and a staggered
potential which can be driven from an ionic to a Mott insulating
state.  In the Mott regime the presence of a ``vacuum'' allows us to
calculate  the different contribution to the gap and to highlight
an ultranonlocality of BALDA. 
\end{abstract}
\date{\today}
 \maketitle

\section{Introduction}
The extension of Density Functional Theory (DFT) to treat lattice
fermions  dates back to the 80's and it has been recently the subject of a revived interest. \cite{kohn1983,chayes1985,gunnarsson1986,sandoval2003,sandoval2004,Saubanere2009Scaling,schonhammer1995,schindlmayr1995,lima2002,lima2003,lorenzana2012,stoudenmire2011,brosco2013,tokatly2011,tokatly2012,akande2012,franca2012,schenk2008,xianlong2012,karlsson2011,campo2007,xianlong2006,verdozzi2008,kurth2010,stefanucci2011,evers2011,troster2012,bergfield2012,capelle2013,meisner2010,vettchinkina,kartsev2013,xianlong2007}
One motivation to  develop Lattice DFT
(LDFT) is that it provides a ``sand-box'' 
environment where one can study the subtleties of DFT itself, clarify the origin
of inaccuracies in approximate functionals\cite{kohn1983,chayes1985,gunnarsson1986,sandoval2003,sandoval2004,Saubanere2009Scaling,schonhammer1995,schindlmayr1995,lima2002,lima2003,lorenzana2012,stoudenmire2011,brosco2013}
and test ideas on new
functionals.\cite{lorenzana2012} 
Another  motivation is provided by the problem of solving lattice models in the
presence of an inhomogeneous potential.
Lattice models are at the basis of our
 understanding of the phenomenology of strongly correlated,  magnetic
 and disordered systems. Their fundamental
relevance has in recent years motivated  a number of successful
experiments with ultracold atomic gases  in optical lattices
\cite{bloch2008,esslinger2010,bloch2012} fueling
at the same time  the development and refinement of efficient
theoretical tools (see {\sl e.g.}[\onlinecite{lewenstein2007}]) among
which LDFT has become particularly useful.\cite{campo2007,scarola2009,fuchs2011}
Static and time-dependent \cite{tokatly2011,tokatly2012,akande2012} lattice DFT were indeed used to investigate the physics of Hubbard models with on-site interaction \cite{franca2012,schenk2008,xianlong2012,karlsson2011,brosco2013,campo2007,xianlong2006}, Kondo models \cite{verdozzi2008,evers2011,kurth2010,stefanucci2011,troster2012,bergfield2012}, disordered interacting lattice models \cite{vettchinkina,kartsev2013} and  spin liquids \cite{xianlong2007}.

The Local
Density Approximation (LDA) is the commonest approximation to the
exchange-correlation (xc) functional of DFT,  already proposed by Hohenberg and  Kohn \cite{hohenberg}, it was first applied to a lattice model by
Gunnarsson and Sch\"onhammer \cite{gunnarsson1986,schonhammer1995}
and subsequently by Lima et al.\cite{lima2003} In the 
case of the one-dimensional Hubbard model there is the advantage that
the exact solution of the homogeneous reference model is known by
Bethe {\sl Ansatz}\cite{lieb1968} (BA). One can thus base the LDA on the 
exact energy or use the approximate but accurate analytical
expressions avaialble\cite{lima2002,lima2003,franca2012}.
One intriguing aspect  of such Bethe {\sl Ansatz} LDA 
(BALDA)  functional
is that, differently from the standard continuum LDA, it has a
discontinuous functional  derivative.
Such non-analytic behavior stems directly from electron correlation
and it has important consequences on the relation between the charge
gap and the gap in the Kohn-Sham (KS)
spectrum. \cite{perdew1982,perdew1983,sham1983,sham1985,godby1986,sham1988,gruning2006,cohen2008,yang2012,brosco2013}
Furthermore, as it will be explained below, it is ultimately
responsible for an ultranonlocality of BALDA, which is absent in
standard continuum LDA. 

Derivative discontinuities and the  associated cusp singularities
in the exchange-correlation energy  undermine the convergence of
self-consistent KS equations. 
So far various approaches have been proposed to solve this problem for the BALDA functional including smoothening of the cusp minimum by going to  a finite temperature \cite{xianlong2012} or by adding an {\sl ad hoc} parameter\cite{kurth2010,karlsson2011} or by relying on Thomas-Fermi approximation\cite{xianlong2006}.
 Here we present a solution which allows us to treat in a simple and
 clean  manner the cusp singularity of the BALDA functional at zero
 temperature: instead of the site occupation we use the LDA local
 chemical potential as a variable and we develop
 self-consistent equations.  
As an example,  we apply our method, which we call $\mu$-BALDA, to  a
Hubbard model subject to a staggered spin-independent  site potential,
also known as the ionic Hubbard 
model\cite{nagaosa1986,zaanen1985,balseiro1989,lima2003} (IHM).  At
half-filling, by modifying the ratio between Hubbard interaction and
the staggered potential it can be tuned continuously from an ionic 
to a Mott insulating regime \cite{nagaosa1986,manmana2004}. In both regimes we calculate the {\rm xc} potential and the charge gap by applying $\mu$-BALDA.

The paper is organized as follows. In Section~\ref{model} we give a  description of the
model and of the basics of  Lattice DFT.  In Section~\ref{BALDA} we present a brief review of BALDA, we explain how the
cusp problem emerges and we  introduce  the  $\mu$-BALDA algorithm. We then apply our method to the 
IHM  in Section~\ref{application}. As a proxy to a solid with a
surface we study a bulk system with high binding energy attached to a 
 zero binding energy chain representing the vacuum.\cite{brosco2013}
This geometry allows to have a well defined Kohn-Sham
 potential at all sites, even  those with integer density, and to 
highlight  an ultra-nonlocal behavior of LDA in the
lattice, computing also the  different contributions
 to the charge gap in the  discontinuous situation.  
We conclude in Sec.~\ref{conclusion}.

\section{Lattice Density Functional Theory}
\label{model}
Let us start by outlining of the basics of LDFT. We consider a Hubbard
chain in an inhomogeneous static field $v_x$:
\bea H&=&-t\sum_{x\sigma }(c_{x\sigma }^{+}c_{x+1\sigma
}-n_{x\sigma
}+{\rm H.c.})+U\sum_{x}n_{x\uparrow }n_{x\downarrow
}\nonumber\\ & &+\sum_{x\sigma }v_xn_{x\sigma }
 \label{ham}
\eea
where $c_{x\sigma }^{\dagger}$ creates an electron with spin
$\sigma =\uparrow ,\downarrow $ at site $x$ and $n_{x\sigma
}=c_{x\sigma }^{\dagger}c_{x\sigma }$, while $U$ and $t$ are
respectively the interaction constant and the hopping amplitude.
Notice that we have included an on-site contribution in the definition of the
``kinetic energy'' i.e. in the first term  on the r.h.s. of
Eq.~\eqref{ham}.  
With this choice and ``external potential'', $v_x=0$, the lowest energy
one-particle state has 
zero energy in analogy with the continuum model, which will be 
useful below to simulate a ``vacuum'' region (Sect. \ref{application}).

As in standard Kohn-Sham continuum DFT,\cite{kohn} also in lattice DFT, 
the total ground-state energy can be obtained by minimizing a functional 
written as the sum of three density dependent terms
\begin{equation}
E=T_{\rm KS}+E_{\rm Hxc}+E_v, \label{E}
\end{equation}
where $T_{{\rm KS}}$ is Kohn-Sham
 kinetic energy functional, $E_{\rm Hxc}$ is the
Hartree-exchange-correlation (Hxc) functional and 
$$E_v=\sum_x v_x\rho_x,$$
 with $\rho_x$ denoting the density at site $x$. Notice that the
 ``functional'' is actually a multivariable function of the on-site
 densities $\rho_x$. The functional 
$E_{\rm Hxc}$ is ``universal'' in that it does not depend on the  external potential $v_x$. 

Minimization of the functional with the constraint $\sum_x
\rho_x=N$ leads to the KS equations,
\begin{equation}
[\hat t+v_x^{s}]\varphi _{\kappa,x}=\varepsilon^N
_\kappa\varphi _{\kappa,x}, \label{EqKS}
\end{equation}
where $\hat t$ denotes the hopping operator $\hat t\varphi
_{\kappa,x}=-t(\varphi _{\kappa,x-1}+\varphi _{\kappa,x+1}-2\varphi _{\kappa,x})$ and $\varepsilon^N_\kappa$ 
indicates the $\kappa$-th eigenvalue of the $N$-particle
system and $\kappa = k\sigma$ includes the orbital $k$ and spin
components. The Kohn-Sham potential is defined as, 
\begin{equation}
  \label{eq:kspot}
v_x^{s}= v_x^{{\rm Hxc}}+v_x.
\end{equation}
Even if not explicitly indicated $v_x^{s}$ and $v_x^{{\rm Hxc}}$
depend on the number of particles $N$. 
In the lattice formulation the functional derivative with respect to
the density becomes a partial derivative with respect to the on-site
density\cite{chayes1985,capelle2013} leading to the following
definition of the Hxc potential, 
\begin{equation}
  \label{eq:vhxc}
v_x^{{\rm Hxc}}=\frac{\partial E_{{\rm Hxc}}}{\partial \rho _x}.  
\end{equation}
The ground state
 density of $N$ particles is composed of all occupied KS
orbitals,
\begin{equation}
 \rho _x=\sum_{\kappa \in 
\text{{\rm occ.}}}\varphi _{\kappa,x}^{*}\varphi _{\kappa,x},
\label{roks}.   
\end{equation}
Since the $v_x^{{\rm Hxc}}$ is a functional of the total density, 
Eqs.(\ref{EqKS},\ref{roks}) have to be solved self-consistently.

Due to the constraint on the total number of particles the  Hxc
potential is defined up to a constant both in the continuum and in the
lattice.  One can extend DFT by considering  ensemble
densities\cite{perdew1982,cohen2008,morisanchez2008,yang2012}. In this
case even the constant term in $v_x^{{\rm Hxc}}$ is determined. 

Intrinsic to the ensemble formulation of DFT are derivatives
discontinuities of the 
exchange-correlation energy functional  when the total  density
crosses an integer $N$. As first discussed by Perdew {\sl et al.}
\cite{perdew1982} these lead to  a discontinuous uniform 
change, $\Delta_{xc}$, in the KS potential  $v_x^s $
when an integer filling is approached from the left or from the right, namely,
\begin{equation}
  \label{eq:deltaxc}
  \Delta_{xc}=v_x^{{\rm Hxc}}(N^+)-v_x^{{\rm Hxc}}(N^-)
\end{equation}
where $N^{\pm}=N{\pm}\eta$ with $\eta=0^+$.
It can be shown that the charge gap of the system
$\Delta_c\equiv E_0(N+1)+E_0(N-1)-2E_0(N)$, with  $E_0(N)$ the
$N$-particle ground state energy,  
satisfies, 
%
\begin{equation}
  \label{eq:chargegap}
\Delta_c=\Delta_{KS}+\Delta_{xc}
\end{equation}
where $\Delta_{KS}=\varepsilon^N
_{N+1}-\varepsilon^N_{N}$  is the single particle Kohn-Sham gap. 


Another important result is the DFT version of Koopman's theorem 
 which is valid both in the
 continuum\cite{perdew1983,Almbladh1984Densityfunctional} and in the
 lattice\cite{brosco2013} and  
it relates the highest occupied Kohn-Sham  eigenvalue $\varepsilon^N_N$
to the ionization energy, 
\begin{eqnarray}
  \label{eq:koopmans}
I_N=-\varepsilon^N_N,
\end{eqnarray}
where  $I_N\equiv E_0(N-1)-E_0(N)$.

One can also show\cite{levy1984} 
that if particles are bound in a finite region of
space around the origin, 
$v_x^{{\rm Hxc}}\rightarrow 0 $ when $x \rightarrow \infty$, such that
  $\rho_x \rightarrow 0$.  

\section{Bethe Ansatz Local density approximation}
\label{BALDA}

Within BALDA   the Hxc energy functional is approximated by a sum of
local contributions as follows\cite 
{kohn1999,lima2003,lima2002}
\begin{equation}
E_{{\rm Hxc}}=\sum_xe_x^{{\rm Hxc}}=\sum_x[e^{\hom }(U,\rho
_x)-e^{\hom }(0,\rho _x)],  \label{Exc}
\end{equation}
where $e^{\hom }(U,\rho )$ is the energy {\sl per}-site  of the 
standard Hubbard model
defined as,
\begin{equation}
  \label{eq:hub}
H_H=-t\sum_{x\sigma }(c_{x\sigma }^{+}c_{x+1\sigma}+{\rm H.c.})+U\sum_{x}n_{x\uparrow }n_{x\downarrow}.
\end{equation}
We do not include here the on-site term in the kinetic energy
which does not affect Eq.~\eqref{Exc} but it affects the zero of the homogeneous
chemical potential defined below. 

In the one-dimensional case the energy of the uniform system can be calculated exactly  for all fillings by BA  \cite{lieb1968} and  one can easily relate the appearance of a finite $\Delta_{xc}$
to the presence of a cusp
singularity in the BA energy density, $e^{\hom }(U,\rho )$, at $\rho=1$. 
The physical  consequences of this non-analytic behavior and the solutions of the related technical difficulties in the implementation of KS-DFT are the subject of the following sections.
Most of the
results presented below can be generalized to higher dimension using 
numerical solutions

\begin{figure}
\begin{center}
\includegraphics[width=0.48\textwidth]{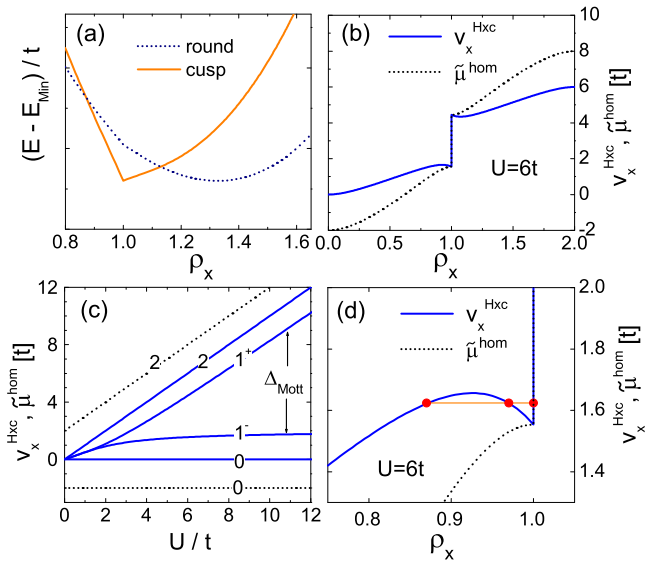}
\end{center}
\caption{ (color online). (a) Examples of cusp and round minima in
BALDA  for $U=6t$ (cusp) $U=3t$ (round)  in the presence of an ionic
potential of amplitude $V=1 t$. In the presence of an inhomogeneous
potential both situations coexist on the same system.    
 (b)Hxc and chemical potentials, $v_x^{\rm Hxc
}$ and $\tilde \mu _x$,  as functions of the site occupation  for $U=6t$
and $v_x=0$.  (c)$v_x^{\rm Hxc
}$  (full line) and $\tilde \mu _x$ (dotted line) as functions of the
interaction strength. Curves are labeled by the on-site occupation
$\rho_x$. In the case of $\rho_x=1^{\pm}$ the two quantities coincide
(only $v_x^{\rm Hxc}$   shown). 
 (d) A zoom of panel (b) around
$\rho _x=1$ to compare the non-monotonous behavior of  $v_x^{\rm Hxc
}$ with the monotonous behavior of $\tilde \mu _x$. \label{fig-cusp}}
\end{figure}
%

\subsection{ Cusp problem and $\mu$-Behte-Ansatz Local-Density Approximation }
\label{metod}
Most difficulties in the implementation of BALDA are related to
the fact that BA energy, $e^{\hom }(U,\rho)$ has a cusp at
$\rho=1$. The implications for BALDA functional are evident
in Fig.~\ref{fig-cusp}(a) where we show two energy curves
describing the typical dependence  of  BALDA energy on the
occupation, $\rho_x$, of a site: when the energy minimum with
respect to $\rho_x$ is located away  or at $\rho _x=1$, the
functional  has  respectively a round  or a cusp behavior at
equilibrium.  In the latter case, the Hxc potential is a
discontinuous functional of the density and, as recently discussed
in  Ref. [\onlinecite{xianlong2012}], the convergence of KS
self-consistent cycle is not guaranteed. 
More precisely, when the density of a site is away from half-filling,
$\rho_x \neq1$, the Hxc potential is obtained 
as usual as the derivative of the Hxc-energy [Eqs.~\eqref{eq:vhxc},~\eqref{Exc}],
\begin{equation}
v_x^{{\rm Hxc}}=\tilde\mu^{\rm hom}(U,\rho _x)-\tilde\mu^{\rm hom}(0,\rho _x).  \label{Vxc}
\end{equation}
where the local chemical
potential, $\tilde\mu^{\rm hom}(U,\rho)$, coincides with the chemical potential of a homogeneous system with density, $\rho$ and interaction $U$,
\begin{equation}
\tilde\mu^{{\rm hom}}(U,\rho _x )=\frac{\partial e^{\hom }}{\partial
\rho _x}(U,\rho _x ) .\label{muhom}
\end{equation}
\begin{figure}[t]
\begin{center}
\includegraphics[width=0.4\textwidth]{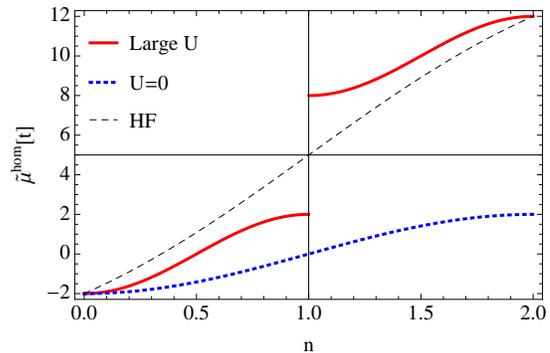}
\end{center}
\caption{Uniform chemical potential for large $U$ as a function of filling. According to Ogata and Shiba\cite{oga90} in the $U/t\rightarrow \infty$ the system can be mapped to a spin-less fermion model. Therefore the chemical potential for $0<n<1$ has a simple sine behavior. At $ n=1$ the chemical potential jumps by the Mott gap. In the picture we took $U=10$ to have a finite gap. For $n>1$ the chemical potential has again a sine form.  For comparison we also plot the chemical potential of non-interacting electrons and the HF chemical potential.} \label{fig-chem}
\end{figure}

On the contrary, for $\rho_x=1$,  the standard definition given in
Eq. \eqref{Vxc} becomes ambiguous, indeed
the derivative of  $e^{\hom }(U,\rho)$ has a jump whose amplitude equals the Mott
gap $\Delta_{\rm Mott}^{\rm hom}$ of the uniform system.  
Moreover, Eqs.\eqref{Vxc}  and \eqref{muhom} 
 together with the definition of Eq.~\eqref{eq:hub},  
imply that $v^{\rm Hxc}$ and $\tilde\mu^{\rm hom}$ have  the same
limits when $\rho\rightarrow 1^{\pm}$, as shown in Fig.\ref{fig-cusp}(b). We can thus define: $v_{\rm
  Hxc}^{\pm}=\tilde\mu^{\rm hom}_\pm$ and we have  
$\tilde\mu^{\rm hom}_{+}-\tilde\mu^{\rm hom}_{-}=v_{\rm
  Hxc}^{+}-v_{\rm Hxc}^{-}=\Delta_{\rm Mott}^{\rm hom}$. Fig.~\ref{fig-cusp}(c)
shows $v_{\rm Hxc}^{+}$, $v_{\rm Hxc}^{-}$ (full lines labeled
$1^\pm$) as a function of $U/t$.  

Not only one value of the density ($\rho_x=1$) corresponds to a
continuum of values 
of $v_{\rm Hxc}$ but also certain values of $v_{\rm Hxc}$ correspond
to three values of the density as one can see in Fig.
\ref{fig-cusp}(d). This leads to numerical 
 instabilities if one uses a standard approach to solve Kohn-Sham
 equations.

As mentioned in the introduction, various methods have been developed to treat these problems : some
modify the BALDA functional relying on a sort of Thomas-Fermi
approximation \cite{kurth2010,karlsson2011}, others entail the introduction of  an {\sl ad
hoc} parameter \cite{xianlong2006} or a finite temperature to smoothen the
discontinuity\cite{xianlong2012}.
%
%
%
%
Here we propose a different route which allows to solve 
the KS  equations  self-consistently  without any additional
parameter. Instead of $\rho _x$, 
we take the chemical potential of the homogeneous system as a
variable. Since, as shown in Fig. \ref{fig-cusp}b, Eq.~\eqref{muhom} is a monotonous growing function of
$\rho$, it can be inverted. We use $\tilde \mu_x $ as the independent 
variable to avoid confusion with the function $\tilde\mu^{\rm hom}$ thus the
inverse of Eq.~\eqref{muhom} is written as, 
\begin{equation}
\rho_x = \rho
_x^{\hom }(U,\tilde \mu _x).\label{romu}
\end{equation}
Then we express the LDA Hxc potential in terms of $\tilde\mu _x$,
\begin{equation}
 v_x^{{\rm Hxc}}(\tilde \mu_x) =\tilde  \mu _x-\tilde\mu^{\rm hom}(0,\rho _x^{\hom
}(U,\tilde \mu _x)) \label{vhxcmu}.
\end{equation}

%

%

  KS equations with the Hxc
potential given in Eq.\eqref{vhxcmu} define the density as 
an implicit functional of $\tilde \mu$: 
\be
\rho _x=\rho _x[v^{\rm Hxc}(\tilde \mu)],\label{romuimp}
\ee
where the square brackets abbreviate the multivariable functions,
{\sl i.e.} we set $\rho _x(v_1^{\rm Hxc}, ..., v_L^{\rm Hxc})=\rho
_x[v^{\rm Hxc}]$. At self-consistence Eqs. \eqref{romu} and
\eqref{romuimp} should give the same result, we therefore have to
impose
\begin{equation}
F_x[\tilde \mu]\equiv \rho _x[v^{\rm Hxc}(\tilde \mu)]-\rho _x^{\hom }(U,\tilde \mu
_x)=0. \label{Feq0}
\end{equation}
Eqs. \eqref{Feq0} can be solved iteratively starting from some initial guess for the chemical potentials $\mu_x^{(0)}$.
Here we expand $F_x$ to arrive at a set of linear recursive equations:
\begin{equation}
F_x[\tilde \mu^{(r-1)}]+{\boldsymbol \nabla}_{\tilde \mu}
F_{x}[\tilde \mu]|_{\tilde {\boldsymbol \mu}=\tilde {\boldsymbol \mu}^{(r-1)}} \cdot
(\tilde{\boldsymbol \mu}^{(r)}-\tilde {\boldsymbol \mu}^{(r-1)})=0,
\label{iterationEq}
\end{equation}
where  we defined the vector ${\boldsymbol \mu}=(\mu_1, ...\mu_L)$.

For a reasonable initial set $\tilde \mu_x^{(0)}$ Eqs.\eqref{iterationEq}
converge rapidly to the final solution while, in the presence of cusps, the conventional KS
iteration scheme does not converge. We stress that our
algorithm treats the cusp and normal sites in a unified way. One
does not need to assume or guess beforehand which site will be a
cusp. If the self-consistent solution $\tilde \mu_x$ falls inside
(outside) the range $[\tilde\mu^{\rm hom}_+, \tilde\mu^{\rm hom}_-]$, 
the site $x$ is a cusp (normal) site (see
Fig.\ref{fig-MottRegime}(a)). It is thus well-suited to describe systems where compressible and incompressible phases coexist.
The gradient in Eq. \eqref{iterationEq} can be computed analytically using perturbation theory
or numerically from finite differences. We found the latter option to be
faster. 

To understand the results discussed below concerning the inhomogeneous
potential case, it is useful to discuss the behavior of $\tilde \mu^{\rm hom}$ and
$v_{\rm Hxc}$ in the uniform case as a function of interaction strength
and filling.  At zero filling [curves labeled 0 in
Fig.~\ref{fig-cusp}(c)] and independently of interaction strength it is
easy to see that our definitions lead to  
$v_{\rm Hxc}=0$ (solid blue curve)  and  $\tilde \mu^{\rm hom}=-2t$  (dashed
line).  For larger fillings at a given interaction strength,  
$v_{\rm Hxc}$  increases due to the effective repulsion among particles.\\ 
For small $U$, the Hxc potential is dominated by the Hartree part and it
behaves as $v_{\rm Hxc}\approx U \rho_x/2$. This behavior is clear up
to $U/t\sim 2$ from the curves labeled $1^{\pm}$ in
Fig.~\ref{fig-cusp}c which represent $v_{\rm Hxc}^{\pm}$.  For larger $U/t$ the two
curves clearly diverge. Such difference is actually present for any
$U/t$ (but exponentially small for small $U/t$) and represents the
Mott gap.   Furthermore, for large interaction, $v_{\rm Hxc}^{-}$ tends to a constant while $v_{\rm Hxc}^{+}$
 increases linearly with $U$. The limiting value of $v_{\rm Hxc}^{-}$
can be understood from the fact that the charge sector of the 
uniform Hubbard model can be
mapped for large $U$ to a spinless fermion model with bandwidth $4t$
describing the lower Hubbard band for  $\rho_x<1$.\cite{oga90} 
This point is also elucidated by Fig. \ref{fig-chem}. There we show the filling dependence of the chemical potential of a Hubbard chain for large $U/t$,
along with the Hartree-Fock  and the $U=0$ chemical potentials.
We see that, as the filling increases, the ``large $U$'' chemical
potential changes from 
 $-2t$ for zero filling to 
 $2t$ for filling  $\rho=1^-$ while the HF chemical potential increases linearly. Notice that the behavior of the exact BA chemical potential $\tilde \mu^{\rm hom}$ for $U=6t$ shown in Fig.\ref{fig-cusp}b is well approximated by the ``large U'' chemical potential. We also see  that $v_{\rm Hxc}$ has the same
 limiting values as $\tilde \mu^{\rm hom}$ for $\rho \rightarrow 1^{\pm}$. 
For $\rho_x=1^+$ both the chemical potential and $v_{\rm Hxc}$ jump by and
amount equal to the Hubbard gap which is of order $\Delta_{\rm
  Mott}=U-4t$ in this limit. The behavior for filling larger than 1
can be understood using particle hole symmetry.

The fact that $v_{\rm Hxc}(\rho_x<1)$ saturates for large $U$
 (instead of having the naive mean-field behavior $v_{\rm Hxc}=U \rho_x/2$) 
is typical of approaches where correlations are taken into account so
electrons can avoid the large Coulomb cost. The
present picture can be compared with similar results obtained using  Gutzwiller approximation 
(see Fig. 1c of Ref.~[\onlinecite{sei07b}]).
There the self-energy and the uniform chemical potential for
filling $\rho_x<1$ saturate at large $U$ and it jumps by the Mott gap on
passing from  $\rho_x=1^-$ to $\rho_x=1^+$. One can interpret the
limiting value of $v_{\rm Hxc}(\rho_x<1)$ as the effect of a residual
kinematic interaction  between quasiparticles in the lower
Hubbard band.

\begin{figure}[t]
\begin{center}
\includegraphics[width=0.48\textwidth]{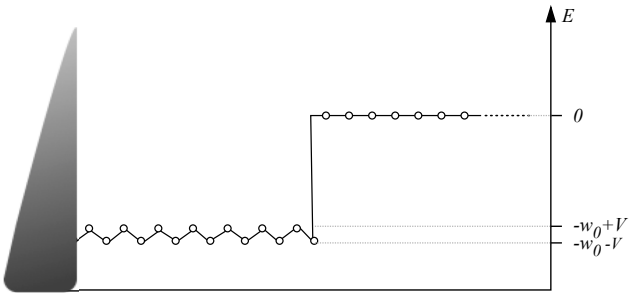}
\end{center}
\caption{Schematic picture of the model studied.} \label{fig-model}
\end{figure}

%
%


\section{Application to the ionic Hubbard model}
\label{application}

We now apply $\mu$-BALDA to the  study of the ionic Hubbard model
(IHM), a Hubbard model with on-site interaction  and a staggered
potential. 


In closed lattice models, the single particle potential that yields the
ground state density of the interacting system is determined up to a
constant since adding a constant to the potential leaves the charges
invariant. In open systems 
which have a ``vacuum'' region one can fix the constant in such a way
that the potential vanishes when both the density and the external
potential vanish, far from the region where the particles
are bound. In Ref.~[\onlinecite{brosco2013}] we considered the open
system shown in
Fig. \ref{fig-model}, with
a vacuum region to completely determine the exact Hxc potential of the ``bulk'' Hubbard
chain from a knowledge of the exact density. 

In BALDA, in principle,  the zero of the Hxc potential is
determined by the explicit  expression \eqref{Vxc}. However,
as explained above, the potential is not determined by this expression 
when $\rho_x=1$ thus to remove any ambiguity on the potential we find
it useful to consider the same geometry as in
Ref.~[\onlinecite{brosco2013}]. 
It will become clear below that in
reality the vacuum region is not essential but it is enough to have  just one
site in  the system  in which $\rho_x\ne 1$ to completely determine
the potential. Still we use the geometry of Ref.~[\onlinecite{brosco2013}]
which is conceptually simple and analogous to the real situation in a
solid. In addition this geometry (or any geometry with a confining
potential) presents a challenge to conventional algorithms because, as we
shall see, cusp sites coexist with non-cusp sites.

  Specifically, we consider a Hubbard chain of
$L_B$ sites with a large binding energy, called ``the bulk'', followed
by a chain of $L_V$ sites with zero binding energy, termed ``the
vacuum'', with open boundary conditions.  The external potential $v_x$
includes both a step-like potential which accounts for the ``work
function''  of the ``solid''  and the staggered field, 
 $$v_x=-[w_0+V (-1)^x]\theta (L_B-x+1/2),$$
 where $w_0>0$ denotes the well-depth and $V>0$ is the amplitude of
 the staggered field. 
 Here we study the model around half-filling and we consider both  the
 band-insulating (BI) and the  Mott-insulating phase (MI)  
the latter appearing when the Hubbbard interaction, 
$U$, dominates
over the staggered potential, $U\gtrsim 2V$ (see {\sl e.g.} Refs.[\onlinecite{nagaosa1986,manmana2004,brosco2013}]).


\begin{figure}[bth]
\begin{center}
\includegraphics[width=0.45\textwidth]{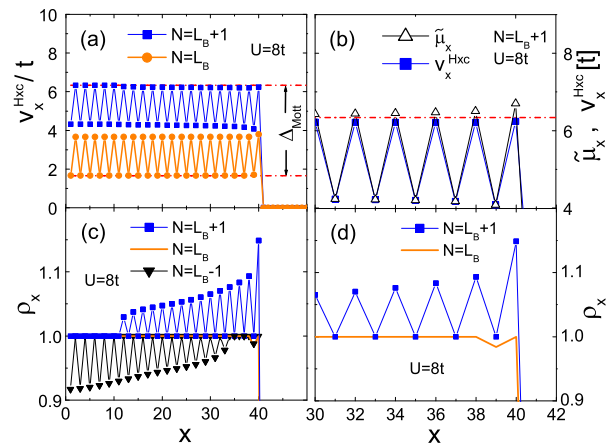}
\end{center}
\caption{(color online).  Panels (a) and (c):  behaviour of the local Hxc potential $v^{Hxc}_x$
and of the density $\rho_x$ for $N=L_B$ and $N=L_B\pm1$ with
$U=8t$, $w_0=10t$, $V=t$ $L_B=40$ and  $L_V=20$.  Panels (b) and (d) are a zoom of panels (a) and (c) where we only plot for $N=L_B$ and $N=L_B+1$ and we also show the local chemical potential $\tilde \mu _x$.
The
dash-dot lines in panels (a) and (b) indicate the potential
$\mu^{+}$ and $\mu^{-}$}. 
\label{fig-MottRegime}
\end{figure}

\subsection{ Shift in the exchange correlation potential  and ultranonlocality of the lattice
  local density approximation} 
Figure~\ref{fig-MottRegime} shows the Hxc potential (panels a and b) and the density (panels c and d) when
the bulk chain is half-filled ($N=L_B$) and when one particle is added
or subtracted respect to half-filling ($N=L_B\pm 1$).   

At half-filling  and for large on-site repulsion the bulk is in the Mott
phase and the system becomes nearly incompressible.  It therefore does not react to the staggered potential and all 
bulk sites have charge $\rho_{x}\sim 1$ (orange line in Fig.~\ref{fig-MottRegime}(c-d)). When this occurs the KS potential becomes nearly
constant in the bulk and  $v^{\rm Hxc}$ acquires a staggered component behaving
as $V (-1)^x$ which  ``screens'' out the external
potential,  as shown in panels (a) and (b) of Fig.~\ref{fig-MottRegime}. This is a strong
correlation effect  captured by BALDA, as opposed to LDA in the
continuum which would not be able to describe a similar situation in  an heteroatomic chain of atoms. 

What determines the value of $v^{\rm Hxc}$ in the bulk 
with respect to the 
vacuum? On a closer look at the density in
Fig.~\ref{fig-MottRegime}(d)   one sees that a small amount of charge leaks to the vacuum from the site with $x= 39$, which is the site with small ionization energy closer to the surface.
Having  a density smaller than 1, this site 
 is not affected by the cusp problem and it has a
well-defined Hxc potential, $v_{39}^{\rm Hxc}\approx v_{\rm
Hxc}^-\approx 2t$. Moreover, since the density has to be homogeneous and all the
sites must have nearly the same effective potential, the Hxc potential of the
other odd sites will be also very close to $v_{\rm
Hxc}^{-}$ while the
Hxc potential of the even sites has to satisfy the relation,
$v_{2x}^{\rm Hxc}=v_{{\rm Hxc}}^{-}+2V$, so that $v^s_{2x}\simeq v^s_{2x+1}\simeq v_{{\rm Hxc}}^{-}+V$.
Notice that all these sites are cusp sites with $\rho_x=1$, 
 so their Hxc potential is not determined by the usual {\em local} 
relation 
$v_{\rm Hxc}=v_{\rm Hxc}(\rho_x)$ but by the ultranonlocal condition that all
sites need to have the same density. Our numerical algorithm
correctly converges to this solution.

For $N=L_B+1$ the
role of odd and even sites is reversed. Indeed, as shown by the
squares in Fig. \ref{fig-MottRegime}(c) in this case, some of the
even sites have a density slightly above 1 and, having their Hxc
potential determined by Eq.~\eqref{Vxc},  they play the role of reference 
sites while the odd sites have all unitary occupations and their
potential is fixed by the even sites potential. 
In this case we have:  $v_{2x}^{\rm Hxc}=v_{{\rm Hxc}}^{+}$ and $v_{2x+1}^{\rm Hxc}=v_{{\rm Hxc}}^{+}-2V$.
The case $N=L_B-1$ is very similar to the case $N=L_B$, in this case,
as explained above, some of the odd sites have a density slightly
below 1 and they play the role of reference sites.

From the above discussion we conclude that the external staggered
potential is screened in the even 
sites for $N=L_B, L_B-1$ and  in the odd sites when $N=L_B+1$.
The net result is that
$v_{\rm Hxc}$ jumps by a constant quantity of order of
$\Delta_{xc}\sim \Delta_{\rm Mott}^{\rm hom}-2V$, with  $\Delta_{\rm Mott}^{\rm hom}\sim U-4t$ and  
where we used the fact that $v_{{\rm Hxc}}^{+}-v_{{\rm
    Hxc}}^{-}=\Delta_{\rm Mott}^{\rm hom}$. The relation with the charge gap
will be discussed in Sec.~\ref{sec:charge-gap}. 

\begin{figure}[tb]
\begin{center}
\includegraphics[width=0.45\textwidth]{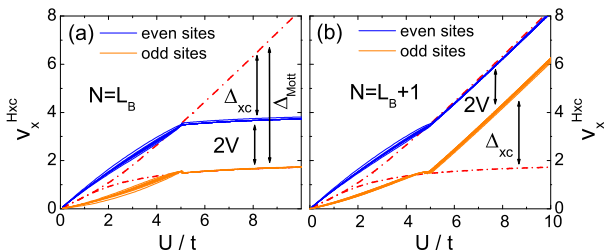}
\end{center}
\caption{(color online). Hxc potential of even (upper blue
band) and odd sites (lower orange band) across the Mott transition
for fillings of $N=L_B$ (panel (a)) and $N=L_B+1$ (panel (b))  for
$L_B=20$, $L_V=20$ and $w_0=15t$. The
dash-dot lines in pane represent
$v_{\rm Hxc}^{+}$ and $v_{\rm Hxc}^{-}$
whose difference equals the Mott gap $\Delta _{\rm Mott}$. }
\label{fig-potentialvsU}
\end{figure}

It is easy to compute the exact ionization energies in the atomic limit
directly from Eq.~\eqref{ham} setting $t=0$. In order to satisfy
Koopmans theorem and in the case of a bulk uniform system (as
found in the Mott regime) the exact KS potential in the bulk 
should satisfy  $v^s_x = -I  $.  Therefore in this limit we can
obtain the exact xc potential quite easily. 
Interestingly, the  KS potentials obtained by $\mu$-BALDA converge to
the same results as shown next. 

At half-filling ionization occurs from odd
sites leading to  $I\simeq w_0-V$ and 
$v^s\sim -w_0+V $ in agreement with the  $\mu$-BALDA results .  
In the case
 $N=L_B+1$ (shown with blue lines in Fig.~\ref{fig-MottRegime}) the
 added charge  will reside mainly on the even sites ($v_x=-w_0-V$)  so
 that  ionization will occur from these sites. In the atomic limit  in
 the Mott regime one 
obtains  $I=w_0+V-U$. This leads to  $v^s\sim
-w_0-V+U $  which implies  $v^{\rm Hxc}_{2x+1}\sim  U -2V \sim v_{\rm Hxc}^+-2V $ and 
$v^{\rm Hxc}_{2x}\sim v_{\rm Hxc}^+ $ again in agreement with the $\mu$-BALDA
results.

Let us conclude the discussion of Fig.~\ref{fig-MottRegime} with a
technical remark. 
 In the exact many-body solution only the uniform compressibility is
zero but the compressibility is small but finite at finite
momentum.Therefore, differently from what predicted by BALDA,  in the exact solution, for all $V\neq 0$  
the charge imbalance between odd and even sites is non-vanishing. 
For large $U/t$, however, the  deviations from the BALDA
$\rho_x=  1$ solution are expected to be small,  the compressibility being trivially zero at all momenta in
the atomic limit.  

Figure~\ref{fig-potentialvsU} 
shows how the xc potential of odd
and even sites evolves from the band insulating regime at small
interaction to the Mott 
insulating regime at large interaction. 
In the band insulator regime the xc
potential tends to screen the external potential 
but it lies outside the interval $[v_{\rm
  Hxc}^-,v_{\rm Hxc}^+]$ marked by the dot-dashed red lines.  In this 
regime the occupation of the sites is larger or smaller than 1. As the
interaction increases the system becomes less compressible and the
charges tend to approach one. When the xc potential of the different
sites  (blue and orange curves)
hit the dot-dashed red lines, (or more precisely the $\mu _x$ fall inside [$\tilde{\mu }_{+}^{\hom },\tilde{\mu }_{-}^{\hom }$]),
the system enters the Mott insulating regime. The density becomes homogeneous
and the difference 
between the odd and even  site potential equals $2V$.
Depending on whether the reference sites have a density smaller 
 or larger than 1, even sites align to $v_{\rm Hxc}^{-}$ (a) or 
 odd sites align to $v_{\rm Hxc}^{+}$ (b).

We stress again that the above results shows how, within BALDA,  strong
correlation leads to a {\sl ultranonlocality} of the Hxc
potential. The potential of the whole system is indeed fixed by  the
presence of just one or  few sites whose density is slightly below or
above half-filling.



As mentioned above, it is enough to have a single site in the system
 with density different from 1 to obtain the absolute value of the xc
 potential in all the sites. Leakage to the vacuum is a
 natural way to obtain such a reference site but an impurity  would work as well.  For a large periodic closed system the
 absolute value of the xc potential can be determined by a limiting
 procedure considering small deviations from half filling from below
 or from above.

\subsection{Charge Gap}
\label{sec:charge-gap}

As stated in Eq.~(\ref{eq:chargegap}) the
fundamental charge gap, $\Delta_c$ can be written as the sum of two
terms, the KS gap, $\Delta_s$, and  a contribution coming from the
discontinuity of $v_{\rm xc}$ upon the addition of an infinitesimal
amount of charge to the system [Eq.~\eqref{eq:deltaxc}].
The latter is site-independent and it can be calculated, in some
simple cases even analytically \cite{gorigiorgi2009,brosco2013}.
A rigorous definition of the discontinuity in  $v_{\rm xc}$
requires using ensemble
DFT\cite{cohen2008,morisanchez2008,yang2012}.  
Alternatively $\Delta_{\rm xc}$ can be estimated employing a formula
first derived by Sham and Schl\"uter in Ref.[\onlinecite{sham1983}],
based on finite differences, which reads 
\be
\Delta_{\rm xc}\simeq \Delta^{N,N+1}_{\rm xc}=\sum_x(v_x^{{\rm xc}}(N+1)-v_x^{\rm
xc}(N))\rho _{N+1,x}^N \label{eq:chargegapSS}
\ee
where $\rho _{N+1}^N$ is the $N+1$'th KS orbital density for $N$
particles and $v_x^{\rm xc}(N)$ denotes the xc potential of the
$N$-particle system at site $x$. 
The difference between $v^{\rm xc}(N)$ and $v^{\rm xc}(N+1)$  on the
r.h.s. of the above equation is in general site-dependent due to the
fact that the charge added is finite. 

\begin{figure}[tb]
\includegraphics[width=0.4\textwidth]{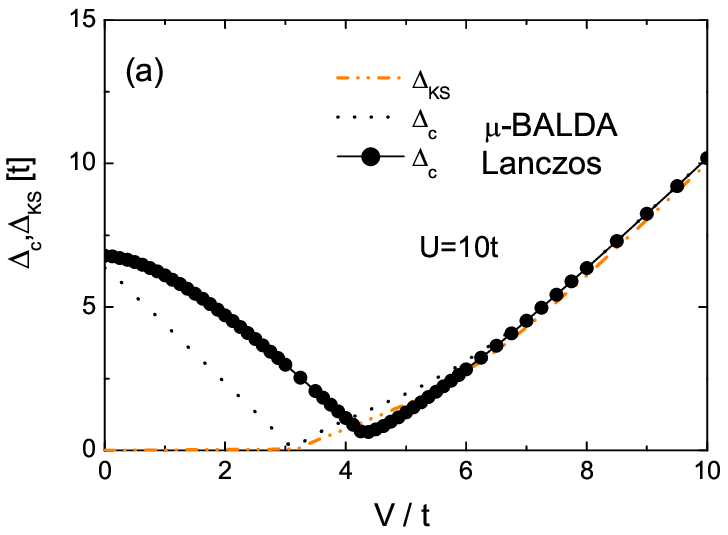}
\includegraphics[width=0.4\textwidth]{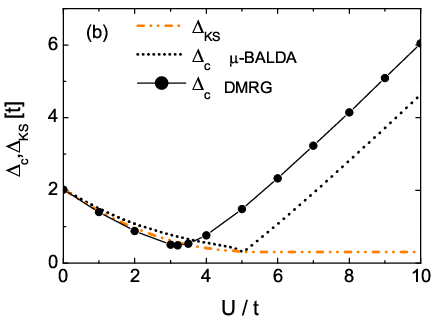}
\caption{(color online). 
(a) Charge gap $\Delta _{c}$ as a function of $V$ for a half-filled
ring of $L=12$ sites with $U=10t$ and periodic boundary conditions. We
  also show exact results obtained with Lanczos exact
  diagonalization\cite{ALPS}. 
(b) Charge gap $\Delta _{c}$ (dashed blue line) and KS gap $\Delta_{KS}$ (dot dashed orange line) calculated with $\mu$-BALDA.
  The results are obtained for  $L_B=20$, $L_V=20$ and $w_0=15t$ and
$V=t$.
For comparison also the  DMRG\cite{ALPS} charge gap is shown (dotted
black line). 
}
\label{fig:chargegap}
\end{figure}

The estimate of  Eq.\eqref{eq:chargegapSS}, which we adopt here,
converges to the exact result of Eq.~\eqref{eq:deltaxc}  when the local
change in the density upon addition of an electron becomes 
infinitesimal small, as in extended systems in the thermodynamic
limit. Eventually we remark that a simple
expression of the error $\Delta^{N,N+1}_{\rm xc}- \Delta_{\rm xc}$
can be obtained along the lines of  the Supplementary Material 
of Ref.~[\onlinecite{brosco2013}] .

We now come to the discussion of $\mu$-BALDA results for the charge gap.
As mentioned in the Introduction, the ionic Hubbard model displays a
transition between an ionic and a Mott insulating regime. As
thoroughly discussed  in a number of papers (see {\sl e.g.}
Refs.~[\onlinecite{manmana2004}] and References therein), the Mott
regime appears when the Hubbard interaction dominates over the
staggered potential and  the charge gap has a minimum at the
transition point.

In Fig. \ref{fig:chargegap} we plot the charge gap (obtained as
total energy differences) as a function of
$V$ (a) and $U$ (b) obtained with  
$\mu$-BALDA and compared to practically exact numerical results.

In the Mott regime the charge gap equals the xc discontinuity and 
the $\mu$-BALDA KS gap  $\Delta_{KS}$ vanishes as the density becomes
homogeneous. On the 
contrary, in the ionic regime the charge gap approximately coincides
with the KS gap while the xc discontinuity becomes vanishingly small, since in
this regime there are no half-filled sites.

Taking into account that the $\mu$-BALDA solution is homogeneous in the Mott
phase we can  give an analytical
expression for the behavior of the $\mu$-BALDA charge gap in the Mott
regime. In this case the charge gap in Eq.~\eqref{eq:chargegap} is
exhausted by the 
discontinuity in the potential obtained in previous Section, 
\begin{equation}
\label{eq:deltac}
\Delta_c  =\Delta_{\rm Mott}^{\rm hom}-2V\simeq U-4t-2V.
\end{equation}
where the last expression is valid for large $U/t$. 
In the large $U$ limit the critical value of the potential is given
by, 
\begin{equation}
\label{eq:vc}
V_c=(U-4t)/2,
\end{equation}
 
 Clearly the transition from the Mott regime
to the band insulating regime occurs when $V\approx 3.0 t$ in $\mu$-BALDA and 
$V\approx 4.3 t$ in Lanczos.

These expressions explain the linear behavior observed in the 
explicit solution of $\mu$-BALDA in the Mott regime [small $V$ in panel
(a) and large $U$ in panel (b)] and also the critical values  
$V_c \approx 3t$ in Fig.~\ref{fig:chargegap}(a) and $U_c=2V+4t\approx
6t$. The latter result overestimates the exact critical $U$ on panel
(b) which we attribute to the inaccuracy of the large $U$
approximation at the critical value. 

In general BALDA underestimates the charge
gap in the Mott regime [small $V$ (a) and large $U$ in (b)]   and it
fails to describe the 
non-linear dependence of the gap for small $V$ (a). These failures can be
related to the fact that in the Mott regime BALDA tends to generate a
ground state which is more homogeneous than the true one as we
remarked in the previous Section. The small inhomogeneity in the exact
density will be associated to a small Kohn-Sham gap in the exact
Kohn-Sham spectrum  which will tend to diminish the discrepancy.  
 For large $V$ in the band insulating regime the $\mu$-BALDA charge
gap rapidly converges to the exact result which approximately
coincides with the KS gap. This shows that in the band insulating
regime, despite the strong interaction, BALDA works remarkably well.

It is interesting to notice that Eqs.~\eqref{eq:deltac},\eqref{eq:vc}
 converges to the exact result in the atomic limit $t=0$. In this case
 the charge gap is given by $\Delta_c\simeq |U-2V|$ and the transition
occurs at $U\simeq 2V$.

\section{  Conclusions and discussions }
\label{conclusion}
Probably one of the most popular LDFT approaches called BALDA 
has till now required {\sl ad hoc} regularizations of the xc energy to
describe the Mott phase. 
In the present work, we have solved this long-standing problem, 
developing a new method to find the exchange-correlation potential of a lattice system
in the Mott regime with a fully self-consistent procedure.
Differently from previous works, our algorithm, which we call $\mu$-BALDA uses the local chemical potentials as variables.
As an example we apply the method
 to the study of the transition between Mott and band insulating regimes of the ionic Hubbard model.
 Beside the general methodological progress,  we obtain several
 results:
(i) we have shown that in the Mott regime the external potential is
completely screened by the xc potential
 (ii) we highlight an ultra-nonlocality of LDA in the lattice, {\sl
   i.e.} we show that in the presence of the discontinuity,  one or
 few ``reference sites'' in the system are capable of fixing the whole xc
 potential; 
 (iii) we calculate separately the different contribution to the gap in
 the different regimes providing,  in particular, an analytical
 understanding of the behavior of the $\mu$-BALDA charge gap in the
 Mott regime. 
iv) In the Mott phase, due to correlation,  ionization  occurs from
 different sites  for a system slightly below or above
 half-filling. This gives a contribution to the ionization energy
 which can not be captured in a single particle picture but which is
 captured by $\mu$-BALDA. Beside these nice qualitative features
  we have discussed the quantitative errors of BALDA in the Mott
  regime and the high-accuracy in the band insulating regime
  even when the interaction is not small.

In this paper we extract the  energy of the homogeneous system from exact BA \cite{lieb1968}, but the method can be clearly also applied when 
the approximate analytical parameterization introduced by Lima et
al. \cite{lima2003,lima2002} is used.
Our results  are therefore relevant
for all 1D lattice models where BALDA is applied, including Kondo
systems\cite{verdozzi2008, stefanucci2011,
evers2011,troster2012,bergfield2012}, dynamical Coulomb blockade
treated with time-dependent DFT\cite{kurth2010}, harmonically
trapped Hubbard electrons
\cite{xianlong2006,xianlong2007,xianlong2012} and spinless
Fermions with neighboring interaction\cite{schenk2008}.
$\mu$-BALDA is actually a rather general approach 
and it could be in principle generalized also to higher dimension or to treat other discontinuous functionals.

\section*{  Acknowlegements  }This work was supported by the Italian
Institute of Technology through the project NEWDFESCM.

\end{document}